\documentclass[a4paper,11pt]{article}
\pdfoutput=1 
\usepackage{jinstpub}
\usepackage{amsmath}
\usepackage{comment}
\usepackage{caption}
\usepackage{subcaption}
\usepackage{url}
\usepackage{hyperref}
\usepackage{xcolor}
\usepackage[section]{placeins}

\title{Argonne Pixel Tracking Telescope at the Fermilab Test Beam Facility}

 \author[a,1]{V. Bhopatkar, \note{Corresponding author. Now at Dept. of Physics, Oklahoma State University}}
 \author[b]{M. Benoit,}
 \author[c, 2]{B. Dong, \note{Now at Dept. of Physics and Astronomy, Michigan State University}}
 \author[d]{M. Kiehn,}
 \author[e]{J. Lambert,}
\author[a]{J. Metcalfe,}
\author[e, 3]{J. Muse,\note{Now at Dept. Physics, University of Minnesota}}
\author[e]{D. Wilbern}
 \affiliation[a]{Argonne National Laboratory}
 \affiliation[b]{Brookhaven National Laboratory}
 \affiliation[c]{Shanghai Jiao Tong University}
 \affiliation[d]{University of Geneva}
\affiliation[e]{University of Oklahoma}

\emailAdd{vbhopatkar@anl.gov, jmetcalfe@anl.gov}

\abstract{The Argonne Pixel Tracking Telescope is installed at the Fermilab Test Beam Facility. The telescope consists of six planar n$^{+}$-in-n silicon sensors with a pixel size of \mbox{250 $\times$ 50 $\mu m^{2}$}. The instrumentation of the telescope is described including the electrical and mechanical setups. A 120~GeV proton beam is used to evaluate the telescope performance using criteria such as cluster size, pixel tracking efficiency, and spatial resolution. The spatial resolution of the telescope is measured to be \mbox{72 $\mu$m $\times$ 13 $\mu$m} and is consistent with the resolution determined from the simulation. This telescope will be used to test the performance of the various silicon pixel technology as well as to study the effects and the performance of the detectors after being irradiated.}


\arxivnumber{2202.05316}

\begin{document}
\maketitle
\flushbottom

\section{Introduction}\label{sec:Intro}
The Large Hadron Collider (LHC) at the Center for European Nuclear Research (CERN) in Geneva, Switzerland, is preparing for an upgrade to the High-Luminosity LHC (HL-LHC). During this period, the HL-LHC is expected to reach a peak instantaneous luminosity of 7.5$\times10^{34} cm^{-2}s^{-1}$, which corresponds to an average of 200 inelastic proton-proton collisions per beam-crossing.

 During the HL-LHC era, ATLAS is expected to receive an integrated luminosity of about 4000 fb$^{-1}$. To operate in this environment, the new tracker detectors should be able to record data with high trigger rates, close to 1 MHz. They also need to have high granularity and a high radiation tolerance. To achieve these requirements, ATLAS is upgrading its Inner Tracker (ITk) Detector, which includes the pixel and strip detectors. New hybrid pixel detectors are under development to adapt to the HL-LHC environment. As in the case of the detectors, the data acquisition (DAQ) system also needs an upgrade to read out upgraded pixel detectors efficiently. These detectors and DAQ systems require intensive testing in a similar environment to ensure they will perform reliably. 

For intensive detector studies, there are dedicated facilities that provide a high-energy beam environment. Such facilities are currently located at CERN and FNAL (Fermilab) among others. For example, the Fermilab Test Beam Facility (FTBF) \cite{FTBF} provides a high-energy (120 GeV) proton beam for testing detectors. A tracking telescope in the test beam facility can provide a testing platform for these different pixel detectors to aid the upgrade efforts. The Argonne Pixel Tracking Telescope (APTT) is permanently installed at the Fermilab Test Beam Facility to aid these testing efforts. This telescope is intended to test the resolution and tracking efficiency of upgraded pixel detectors and to study the impact of irradiation on the performance of these same detectors. The APTT was commissioned using six planar n$^{+}$-in-n silicon sensors \cite{Capeans:1291633}, read out by the FE-I4B~\cite{672298352} front-end chips. The same technology is currently used in the ATLAS Insertable B-Layer (IBL)~\cite{Capeans:1291633} modules. IBL pixel modules were installed in the ATLAS tracker during a past upgrade.

The paper describes the FTBF, APTT instrumentation, DAQ system, reconstruction and simulation software, and the performance of the pixel telescope in detail.

\section{Fermilab Test Beam Facility}
\label{sec:FTBF}
The FTBF is one of a few facilities in the world that provides a high-energy beam with very high intensity.
At the FTBF, the primary beam consists of 120 GeV protons, whereas the secondary beam provides pions, muons, and electrons, with energies ranging up to 66 GeV. The typical beam size is about 1 cm in diameter and can be adjusted to some degree by quadrupole magnets as per the requirement. The beam is delivered in a spill every 60 seconds with a 4.2 seconds duration. The maximum available rate is approximately 2.5 GHz/cm$^{2}$, which is about 5E5 particles per spill.

The APTT is installed in the MTest (MT6) enclosure. The FTBF facility provides scintillator counters with wire chambers at each enclosure to provide the particle count rate. The MT6 enclosure is equipped with patch panels to access the local network from the control room. Also, there is access to inert gases such as nitrogen for the experiment.

\section{Instrumentation}
\label{sec:Instru}
The APTT consists of six modules, divided into two sets of three modules. Telescope modules referred as telescope planes are mounted on the railings with adjustable stages along and transverse to the beamline.  The added mechanical structure is convenient for remotely aligning detectors in the beam from the control room. The HSIO-II/RCE~\cite{hsio} DAQ system records the data from all six telescope planes. The test beam data is reconstructed using the "Proteus"~\cite{proteus} reconstruction software, which produces all performance characteristics plots of offline alignment such as cluster size, pixel efficiencies, and residuals. These test beam data results are compared with the simulation results produced using the "AllPix$^{2}$"~\cite{allpix} simulation software.

\subsection{Telescope Planes}
\label{sec:planes}
Six n$^{+}$-in-n planar silicon sensors are used in the Argonne Pixel Tracking Telescope. Each sensor is 200 $\mu$m thick and 41.4 mm $\times$ 37.5 mm in size. This sensor is read out by four front-end FE-I4B chips.  This chip is built in a 130 nm CMOS process and has an active area of \mbox{20.2 $\times$ 18.8 mm$^{2}$.} It has a pixel matrix of 80 columns on 250 $\mu$m pitch by 336 rows on 50 $\mu$m pitch. 

\begin{figure}[!h]
     \centering
     \begin{subfigure}[b]{0.5\textwidth}
         \centering
         \includegraphics[width=\textwidth, angle=0.5]{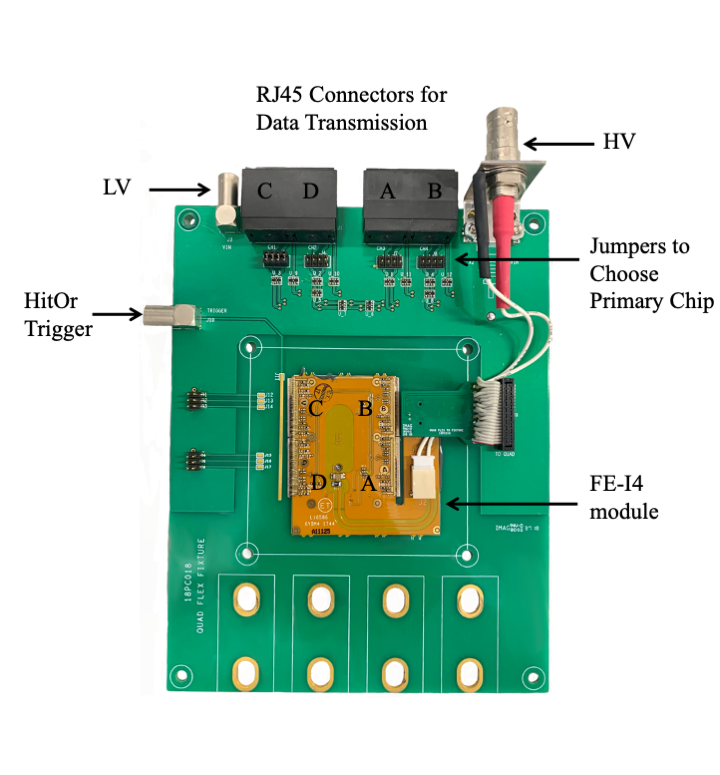}
         \caption{}
         \label{fig:BM}
     \end{subfigure}
     \hfill
      \begin{subfigure}[b]{0.49\textwidth}
         \centering
         \includegraphics[width=0.7\textwidth]{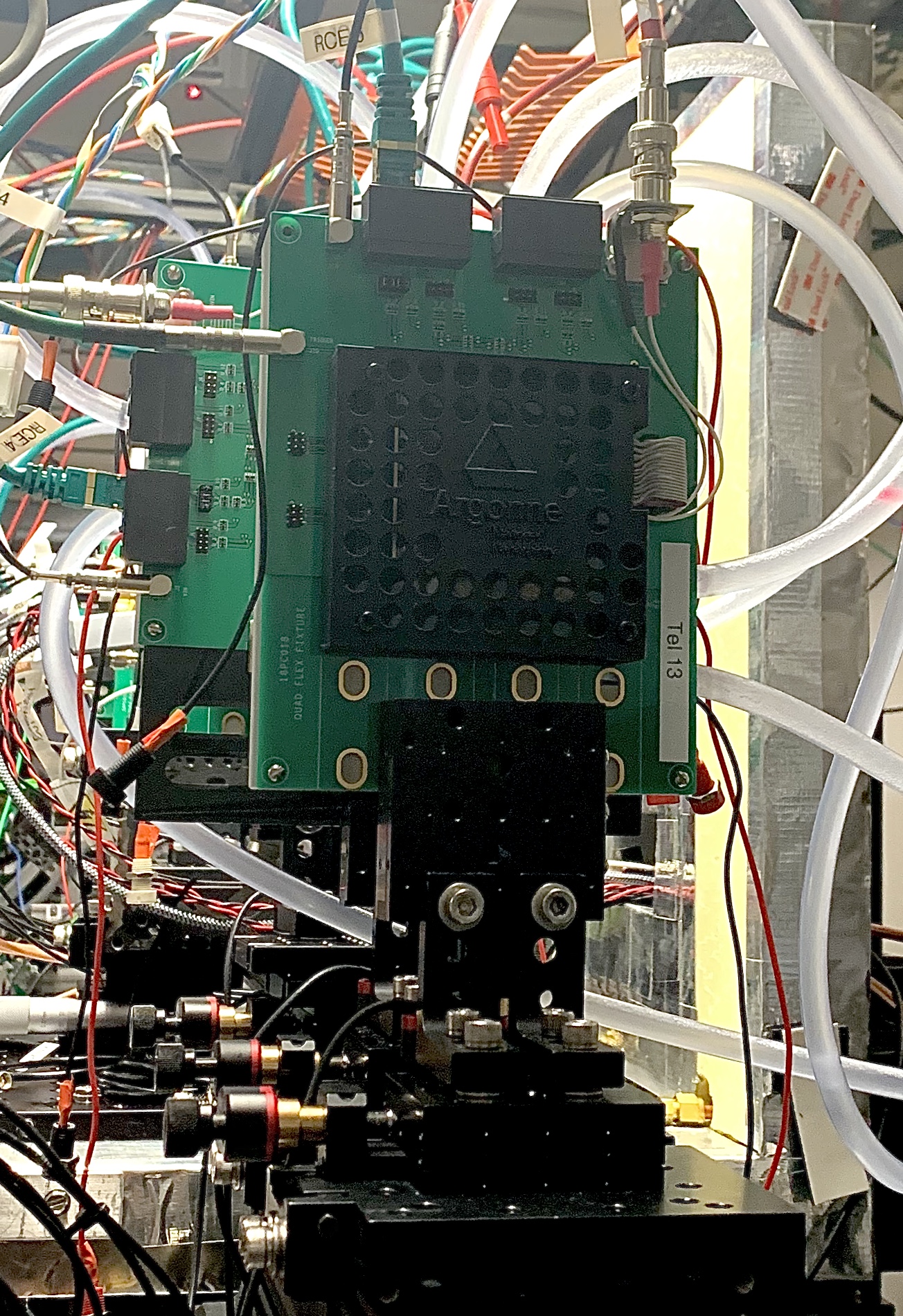}
         \caption{}
         \label{fig:module}
     \end{subfigure}
        \caption{ (a) FE-I4B module glued on the outer PCB. Letters A, B, C, and D indicate four chips on the modules   (b) FE-I4B module mounted on the downstream telescope arm}
        \label{fig:telplane}
\end{figure}
The bare module is a silicon sensor bump-bonded to four FE-I4B chips i.e. each chip is approximately bump-bonded to the 1/4th of the sensor area. During the module assembly, the bare module is glued to the backside of the flexible printed circuit board (flex PCB). Wire bonds between the flex PCB and the chip provide the necessary electrical connection. The assembled module is then glued to the outer PCB board, which is designed for mounting the planes in the beamline. This board consists of RJ45 connections for data links, lemo connections for the low voltage (LV) and internal Hit-OR trigger, and an SHV connection for the high voltage (HV).  Removable jumpers on the data links on the PCB board give the flexibility to choose the primary chip from any of the four chips. Figure \ref{fig:telplane} shows an assembled FE-I4 module on the PCB board and the front view of the module after installing it as a telescope plane in the beamline.

As per Figure \ref{fig:BM}, the 250 $\mu$m pitch is in the X direction and the 50 $\mu$m pitch is in the Y direction. Due to the non-squared shape of the sensor pixels, the second and fourth telescope planes are rotated by 90$^{\circ}$ around the beam axis to achieve comparable spatial resolution in the X and Y direction. Only one chip per module is actively used for data taking. The active chip in each telescope plane is chosen based on its alignment with the beam and other telescope planes and also the constraints on the beam spot size. Table~\ref{table:tab1} summarizes the active chip used for data taking for each telescope plane.
\begin{table}[h!]
\centering
\begin{tabular}{|c|c|c|c|c|c|c|} 
 \hline
 Telescope Planes & Tel0 & Tel1 & Tel2 & Tel3 & Tel4 & Tel5 \\ [0.5ex] 
 \hline
 Active chip & C & B & B & B & D & C \\ 
 \hline
\end{tabular}
\caption{Active chip in each telescope plane for data taking}
\label{table:tab1}
\end{table}

During data taking, the FE-I4B chip measures deposited charge by monitoring a firing of discriminator and time-over-threshold (ToT) with a 4-bit resolution at 40MHz external clock signal. With Digital-to-Analog Converters (DACs), the global threshold on the chip can be set by applying the local correction to each individual pixel. The hit information and 8-bit time stamp are stored in memory cells shared by four adjacent pixels. The chip stores the hit information with a latency interval, which is programmable up to 255 cycles of the external clock. This information is read out when a trigger signal is received in this interval. Synchronized data was collected from the telescope plane by supplying the simultaneous triggers to the first and last telescope planes.

\subsection{Mechanics}
\label{sec:mech}
\begin{figure}[!h]
    \centering
    \begin{subfigure}[b]{0.47\textwidth}
    \centering
    \includegraphics[width=\textwidth]{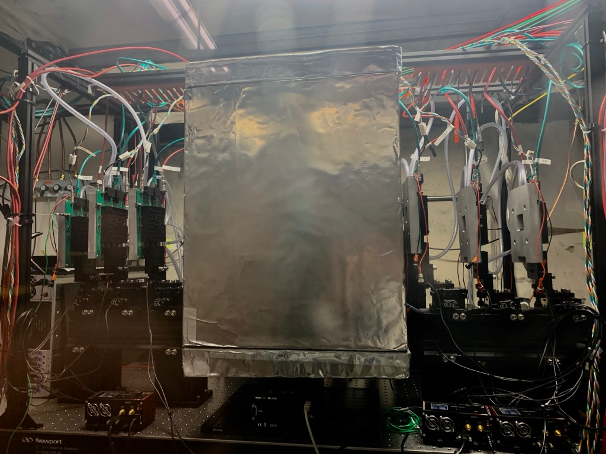}
    \caption{}
    \label{fig:tel}
    \end{subfigure}
    \hfill
    \begin{subfigure}[b]{0.5\textwidth}
    \centering
    \includegraphics[width=\textwidth]{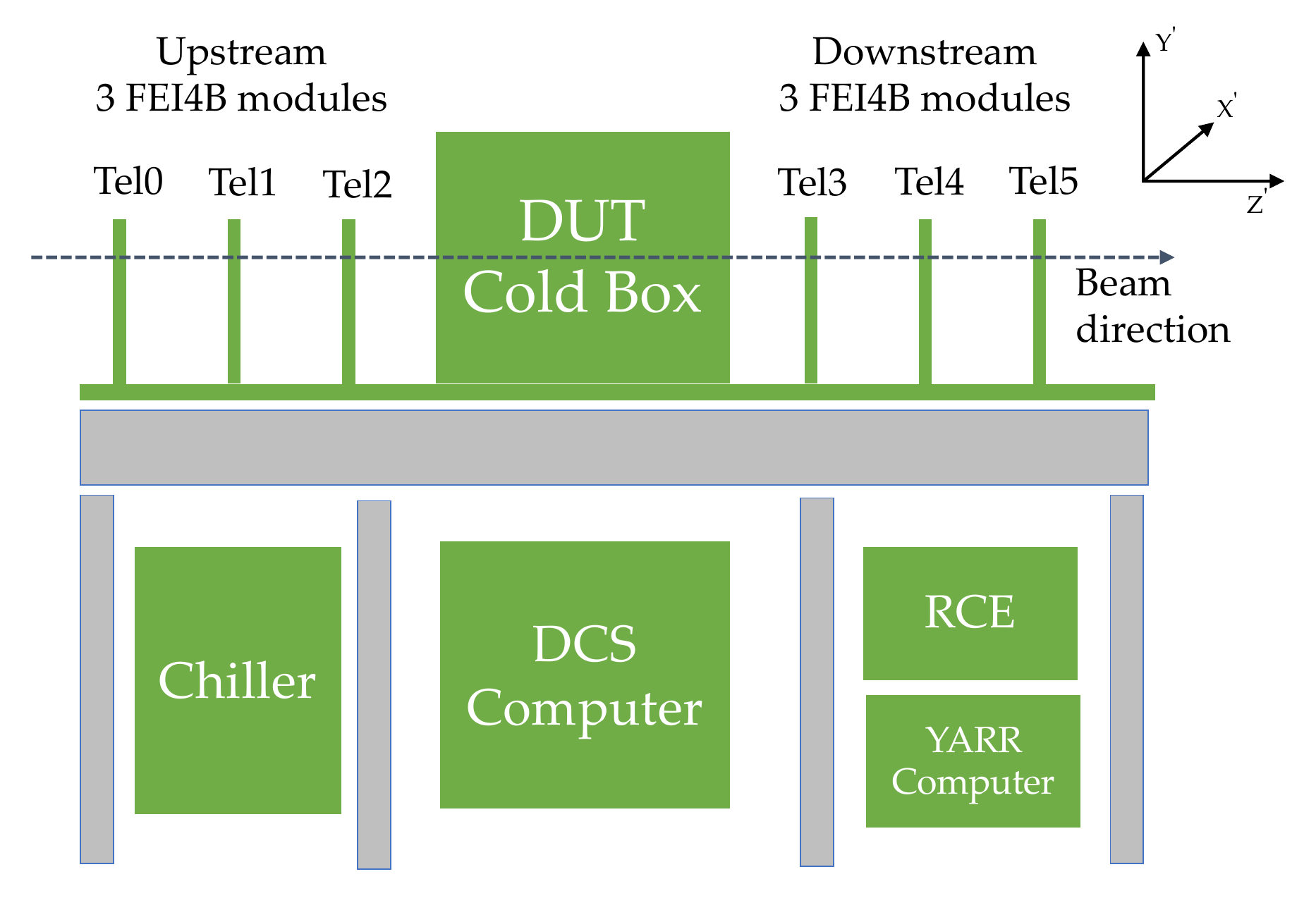}
    \caption{}
    \label{fig:schema}
    \end{subfigure}
    \caption{(a) Telescope setup in the FTBT MT6B enclosure (b) Schematic diagram of the setup }
    \label{fig:schematel}
\end{figure}

The setup of the APTT in the MT6 enclosure at FTBF and the schematic diagram of this telescope are shown in Figure~\ref{fig:schematel}. Telescope planes are installed on rails, referred to as the telescope arms. Each arm consists of three planes, and between these two arms, the cold box is installed on the movable stage. For better alignment, each telescope plane is mounted on a movable stage. The stage can be moved in the X' and Y' directions perpendicular to the beam by 100 mm. These stages are configured to operate remotely.
A cooling system is added for the telescope planes to avoid overheating the chips. This cooling system includes a cold plate with a thermally conductive graphite sheet to transfer heat out of the module for each plane. The cold plate is cooled down to 5$^{\circ}$C by circulating cold distilled water, which helps to maintain the temperature of the unpowered chips around 10$^{\circ}$C and around 30$^{\circ}$C when the chips are powered.

In the test beam setup, a right-handed Cartesian system is used for global and local coordinate systems. In the global coordinate system, the beam points in the Z'-direction and the Y'-direction points vertically upward. In each chip, the lower-left pixel is considered as the origin of the local system. The local frame position determines pixel hits, and then it is transformed into the global frame for tracking purposes. The details about the transformation can be found in Ref. \cite{allpix}.

Racks are mounted under the telescope table to hold electronics, power supplies, data acquisition computers, and chillers. The following subsections include a detailed description of these various types of equipment. These are connected and operated using a local network.
\\
\\
\textbf{Power supply:}
The telescope needs different voltages for operations. Low-voltage (LV) is required to power the chip, and high-voltage (HV) to apply a bias voltage to the sensor (-80V). WIENER MPOD \cite{weiner} is a multi-channel LV and HV computer controlled power supply. It is a 19" mountable chassis with module cages for 10 LV or HV modules. Currently, this MPOD crate is equipped with one iSeg EHS F410n \cite{iseg} HV module and two WIENER OMPV8008 LV modules. The HV module has 16 channels, and each channel can supply up to 1 kV and a maximum of 8 mA current through SHV connectors. The LV modules exhibit 8 channels. The signals from four LV channels are driven through a 37-pin sub-D connector. Local control of the MPOD crate is available using the LCD and currently operating remotely through the local network.
\\
\\
\textbf{PDU and Network Switch:}
The CyberPower PDU15SW8FNET \cite{PDU}, an 8-outlet (front) rackmount switched power distribution unit (PDU) is used, which provides 120V, 15A output. The outlet receptacles can be managed locally or over the network using the PDU console, web browser, and RJ45 Ethernet port for network connection.
For the local network connection, the tp-link 24-Port Gigabit Switch is used. The switch is connected to Fermilab's internal network, which can be accessed from the control room.
\\\\
\textbf{Cooling System:}
The main focus of the pixel telescope is to characterize the Device Under Test (DUT). One of the critical tests for high-energy experiments is to study irradiation effects on the detector. Irradiated modules require cooling to limit the leakage current. DUTs are cooled using the Ultra-Low Refrigerated-Heating Circulator, Julabo FP89-HL \cite{julabo}. Tubing from the chiller is connected to the base plate, which holds the DUTs. For cooling, cold silicon oil is circulated through base plates. The chiller can reach the minimum cooling temperature of -90$^{\circ}$C and it is remotely controlled using a DCS computer via an RS-232 serial socket connection. In addition to the circulating chiller fluid, cooled nitrogen or dry air is available to flush in the cold box to lower the air temperature and maintain the humidity to keep the DUT temperature always below the dew point. This chiller is dedicated for cooling DUT for future studies. With this system the DUT can be cooled down to -35$^{\circ}$C.

The Thermo Fisher Polar Series, Accel 500 LT Cooling/Heating Recirculating Chiller \cite{ThermoFisher} is used to cool the telescope plane. The temperature range for the chiller is from -25$^{\circ}$C to +80$^{\circ}$C. The distilled water is used as a coolant for this chiller.
\\
\\
\textbf{Temperature \& Humidity Sensors:}
It is crucial to monitor the temperature and humidity of the cooled DUT to avoid condensation. Thorlabs TSP01 \cite{web:thorlabs} 2-channel compact USB temperature and humidity logger is used to monitor the DUT system. Logger software GUI provides the graphical interface of the readings. 
\\
\\
\textbf{Cold box for DUTs:}
Irradiated DUTs need cooling for testing; hence a thermally insulated box is installed between the telescope arms as shown in Figure \ref{fig:tel}. Currently, the box can accommodate three DUTs. The box is made up of high-quality foam with an aluminum foil exterior. It is resting on a movable stage and has a sealed port to pass through various cables. The movable stage can be moved manually in the X'-direction with the large area translation plate (TBB1212) in the range of 120 mm, whereas the motorized high load vertical stage (MLJ150) is used in a vertical direction in the range of 50 mm. These stages are from Thorlabs, Inc. \cite{web:thorlabs}. The motorized stage can be controlled via an RS-232 serial port.
A 30 mm $\times$ 30 mm breadboard with embedded cooling tube is installed at the bottom of the box. For cooling the DUT, cold silicon oil is circulated through the breadboard, and cooled nitrogen or dry air is blown in the box. 

This box is not used during characterization studies of the telescope planes but is intended to be used in the future to test the performance of irradiated DUT's.

\subsection{DAQ and Reconstruction Software}
\label{sec:reco}
\begin{flushleft}
\textbf{Data Acquisition System}
\end{flushleft}
The data acquisition system (DAQ) of the pixel telescope consists of a high-speed input-output board (HSIO II) with re-configurable cluster elements (RCE) and a computer with custom designed software for data taking.

The HSIO II is based on the Xilinx Virtex 7 Artix FPGA. It provides specific connectivity to the front-end electronics and processes the data received from it. It also generates the clock and triggers and issues them to all telescope planes. 
The RCE is a generic computational unit based on a System-On-Chip (SOC), which can handle several lanes of high-speed I/O. It configures the HSIO board and also sends commands to the front-ends. The raw data from the HSIO is transferred via Ethernet to the DAQ computer.
The configuration file in the DAQ software allows for setting different global parameters such as chip ID, trigger mode, or delays for each telescope plane. These various settings are also accessible through a graphical user interface (GUI). The software provides online monitoring plots during data taking, including real-time hit maps, correlation plots of the hit position between the neighboring planes, and the charge and time information.
\begin{flushleft}
\textbf{Reconstruction Software}
\end{flushleft}
The test beam data from the telescope planes are saved as raw data in the RCE format. Particle tracks are reconstructed from this raw data using the Proteus reconstruction software. The program starts with the raw hit data and provides a fully reconstructed cluster and track. The software performs reconstruction in three steps, i.e., clustering, alignment, and track reconstruction. For each reconstruction, the global position, orientations of all detector planes, and type of detector for each plane must be provided in configurations using the TOML configuration file format.

RCE data are stored as a ROOT \cite{BRUN199781} n-tuples with a complete event and hit information. During the processing, first noise scans are performed on each telescope plane, which masks all the noisy pixels with noise level 5$\sigma$ above the average. The first telescope plane is considered as the reference plane, and the rest of the tracking planes are coarsely aligned with the reference plane based on cluster correlations. Then the fine alignment is performed using a track-based alignment, which minimizes track residuals. 

The first step towards the event reconstruction is to cluster the pixel hits triggered by the same particle traversing through a sensor plane. Proteus implements a simple recursive clustering algorithm that groups all neighboring hits. Overlapping clusters can not be detected as separate clusters but are considered as one. The actual position of the hit is then calculated as the geometric mean of the pixel positions in that cluster. Next is the track reconstruction, in which the algorithm initially assumes that the track is parallel to the longitudinal direction of the telescope and then modifies the track as clusters are added. For standard operation, the track selection requirement is that all tracks must pass through all six planes. Straight-line fits are then performed on candidate tracks, and the one with the smallest $\chi^2$ is selected. The track position is always influenced by the current alignment of the telescope and therefore, the DUT is always excluded from the track fitting to form unbiased tracks.

\subsection{Simulation Software}
\label{sec:simu}
The simulation studies are performed to cross-check the telescope response regarding the residual widths. Events are simulated using AllPix$^{2}$ (Allpix squared) \cite{SPANNAGEL2018164}, a GEANT4-based simulation software framework written in C++. It provides a flexible interface to create an experimental setup for simulation in any configuration. For each simulation, there are three required layers of configuration files: the detector model configuration, which describes a particular type of detector, the detector configuration, which passes the detector geometry setup to the framework, and the main configuration, which contains both the global framework configuration and the list of modules to instantiate together with their configuration.

The pixel detector model can be defined by the pixel pitch, the thickness of the sensor, the readout chip for a hybrid configuration, the PCB properties, the mechanical support, and other geometry properties related to an assembly. 

In the detector geometry configuration, the model type of all detectors, the number of detector planes, and the detector position and orientation are included. A unique ID number identifies each detector. The detectors' positions in space are specified by the coordinates (x', y', z') and the rotations around the x', y', and z' axes through the center of the detectors.

The main configuration consists of a set of sections specifying the modules used. All installed modules are loaded automatically. \textit{DepositionGeant4} module, an interface to Geant4, deposits charge carriers in the active volumes of all detectors and initializes the physical process to simulate a particle source that will deposit charge carriers for every simulated event. 
An actual test beam setup is simulated with repeated sets of 100k events with the six FE-I4B telescope planes and the central plane on each side rotated by 90$^{\circ}$. The 120 GeV proton beam with a 5 mm radius is assumed. \textit{RCEWriters} module saves the output of the simulation in the RCE format, which is compatible with the Proteus telescope reconstruction software. 
The telescope performance in the test beam and the simulation results are described in the Telescope Performance section~\ref{sec:performance}.

\section{Telescope Performance}
\label{sec:performance}
\begin{figure}[!h]
     \centering
     \begin{subfigure}[b]{0.49\textwidth}
         \centering
         \includegraphics[width=\textwidth]{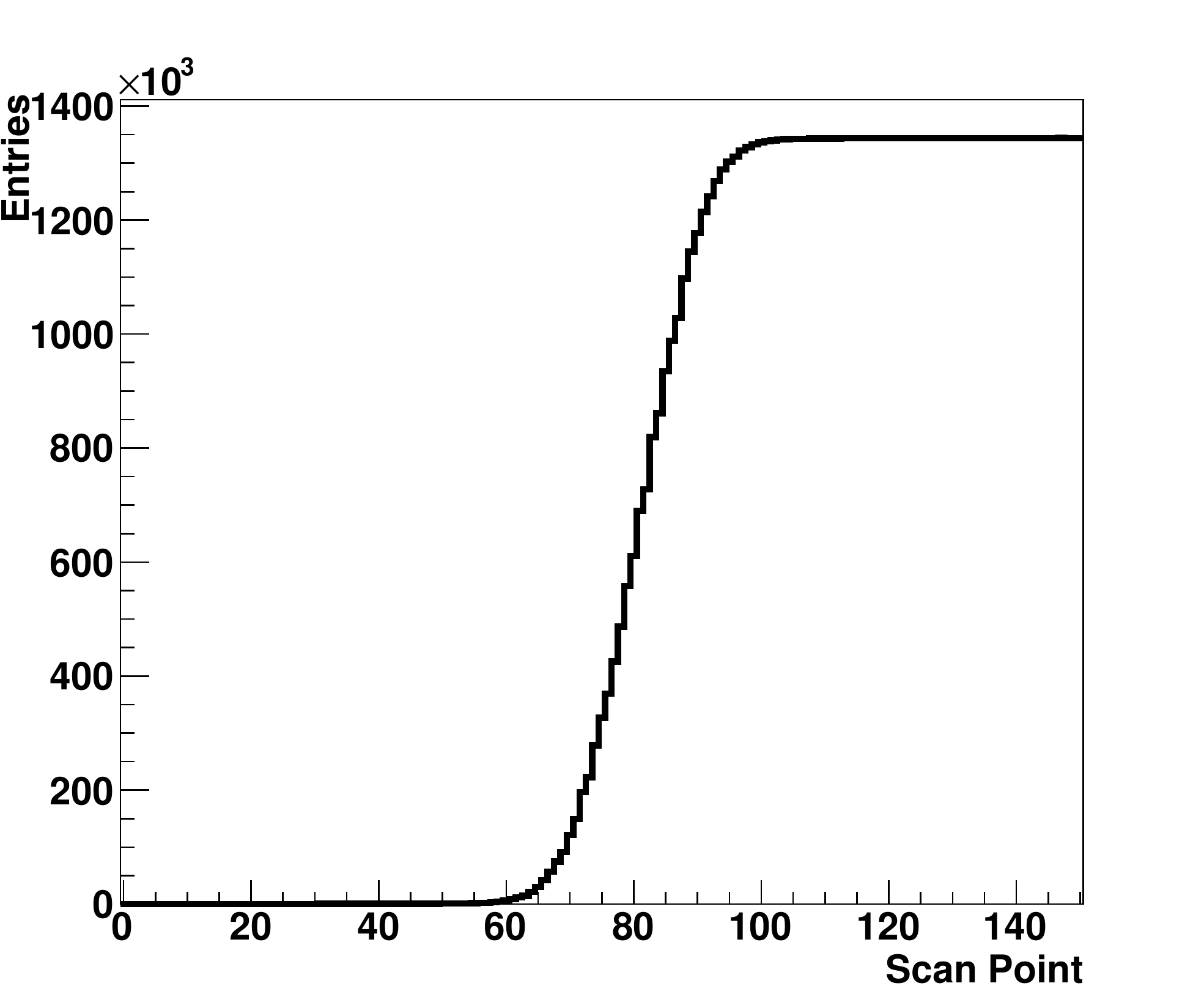}
         \caption{}
         \label{fig:tuneplane1}
     \end{subfigure}
     \hfill
     \begin{subfigure}[b]{0.49\textwidth}
         \centering
         \includegraphics[width=\textwidth]{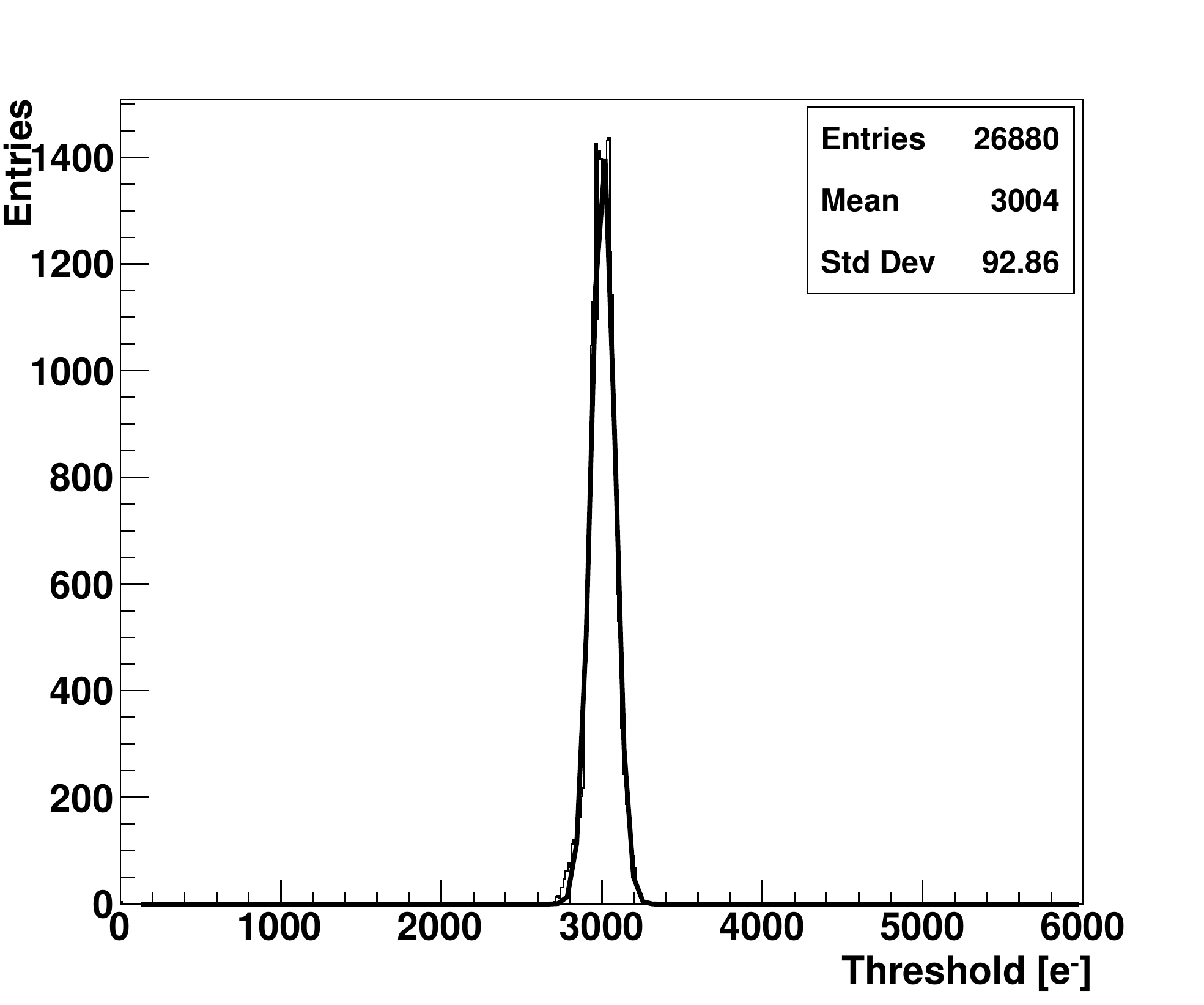}
         \caption{}
         \label{fig:tuneth}
     \end{subfigure}
     \hfill
        \caption{For the first telescope plane (Tel0) (a) S-Curve distribution of the tuned chip  (b) Threshold distribution after tuning the chip}
        \label{fig:tuneplot}
\end{figure}

A custom-made DAQ system is used for data taking at the Fermilab test beam. The DAQ system consists of two GUIs: calibGui and cosmicGui. The calibGui is used for setting the tuning parameters for each chip. Calibration includes various scans through which noisy pixels are masked and the threshold is set. Telescope planes are tuned to a 3000e threshold. The threshold is controlled by the two discriminators located on each pixel front-end chip. The first discriminator controls the threshold for the individual pixel while the second controls an entire front-end chip. In the threshold scan, a fixed charge is injected into the pre-amplifier on the front-end chip. This step is repeated multiple times and the number of hits is recorded as a percentage of injection. Again this step is performed by changing the discrete voltage steps in a pre-amplifier for a specified charge range. Ideally, this process will produce a step function: zero injection resulting in a hit for any charge below the threshold and all injections resulting in a hit for any charge above the threshold. Due to the electronic noise, the curve is smeared in an S-shaped and hence called an "S-Curve". The mean value of the Gaussian error fit function will record the threshold for each function and the sigma value will define the noise of each pixel. Figure \ref{fig:tuneplot} shows the S-Curve and threshold distribution of the tuned chip.

The FE-I4B chip has a self-triggered mode, which can be used to test the detector using an external source before placing it in the beamline.

The cosmicGui is used for real-time data taking. For the data taking, the first and the last planes are used as trigger planes, where HitOr signals from these planes are fed to RCE, and for the given trigger interval, the data is read from all telescope planes. 

\subsection{Occupancy}
\begin{figure}[!h]
    \centering
    \begin{subfigure}[b]{0.45\textwidth}
         \centering
         \includegraphics[width=\textwidth]{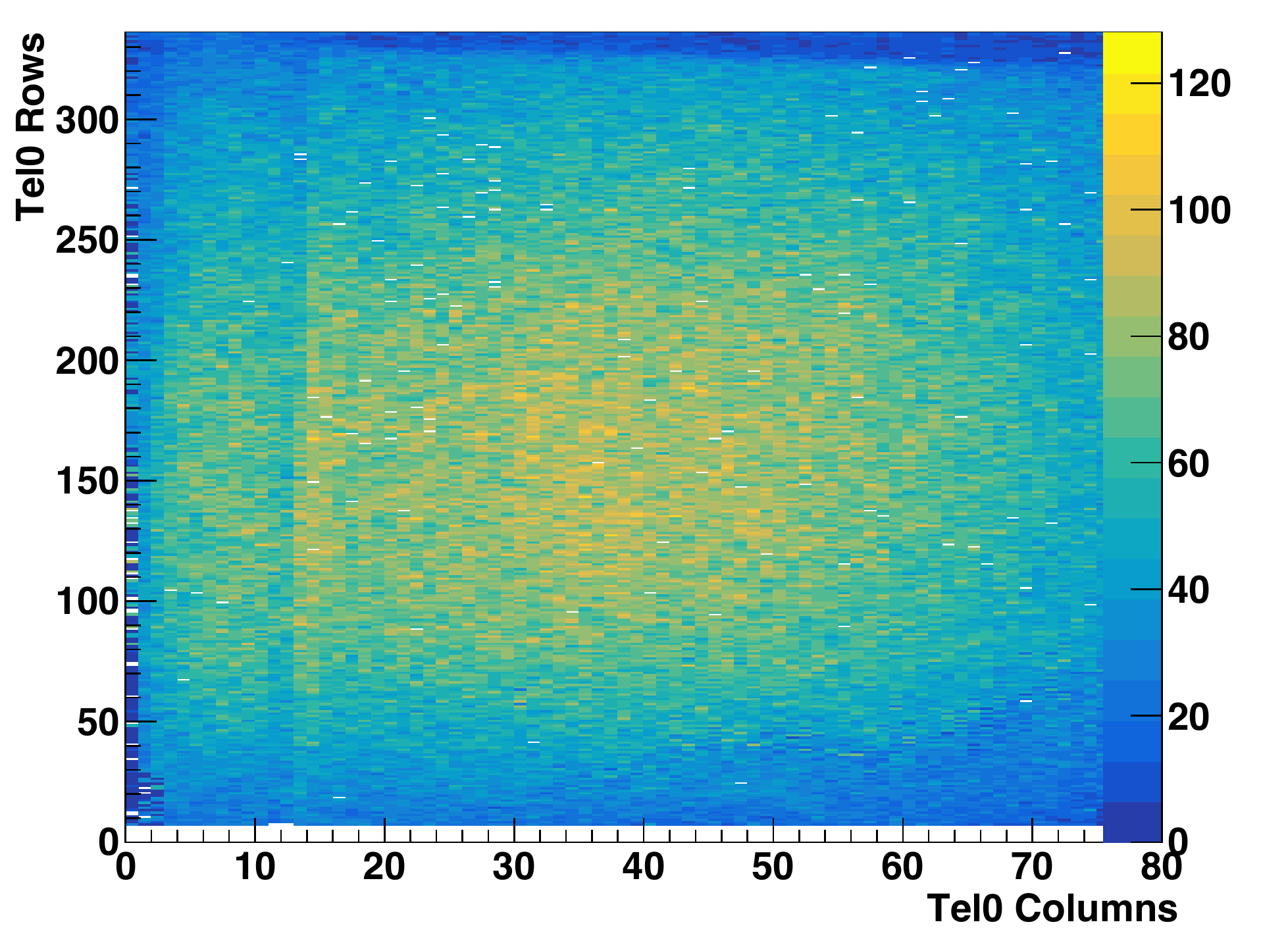}
         \caption{}
         \label{fig:WB}
     \end{subfigure}
     \hfill
    \begin{subfigure}[b]{0.49\textwidth}
         \centering
         \includegraphics[width=\textwidth]{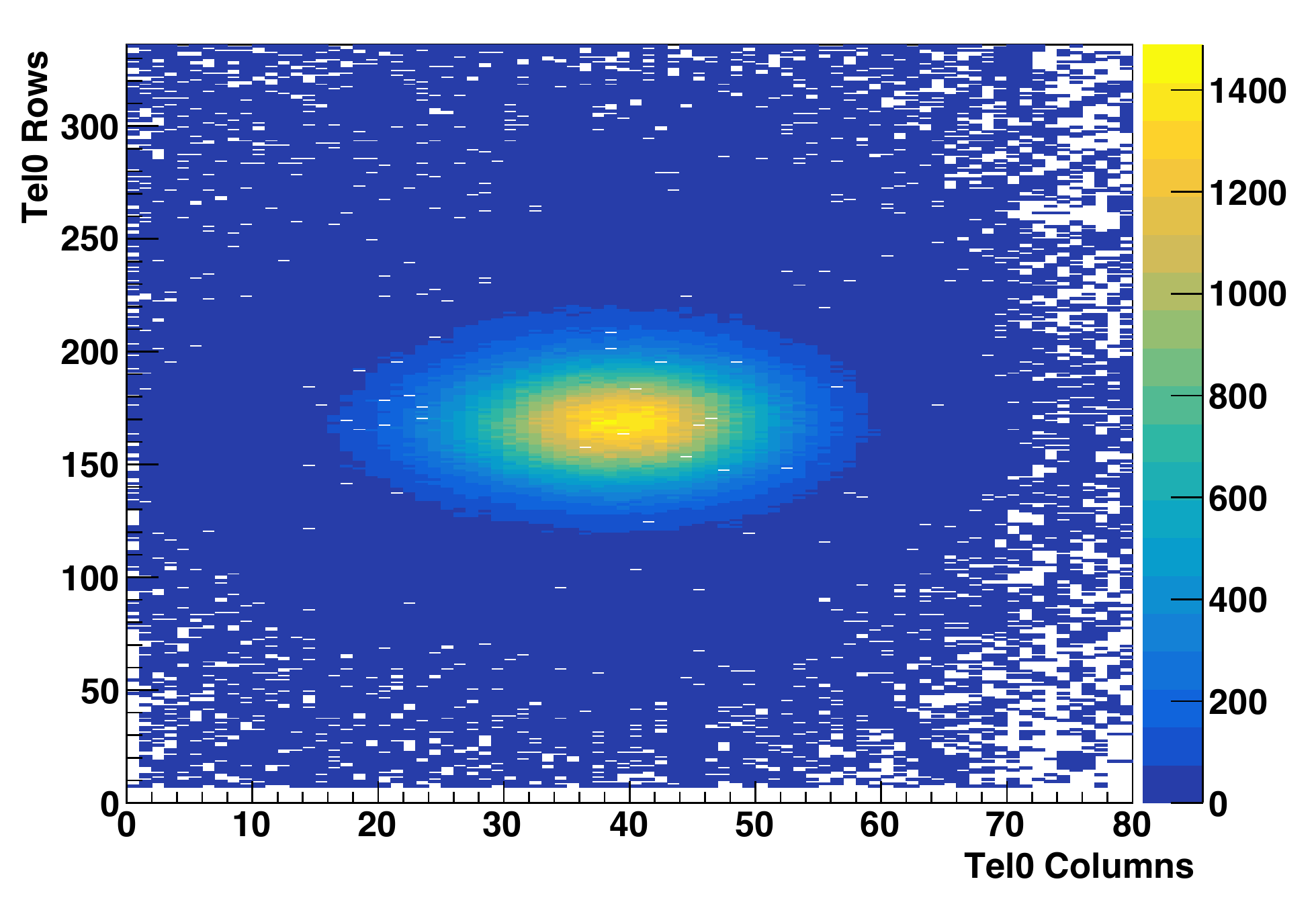}
         \caption{}
         \label{fig:NB}
     \end{subfigure}
     \hfill
     \begin{subfigure}[b]{0.45\textwidth}
         \centering
         \includegraphics[width=\textwidth]{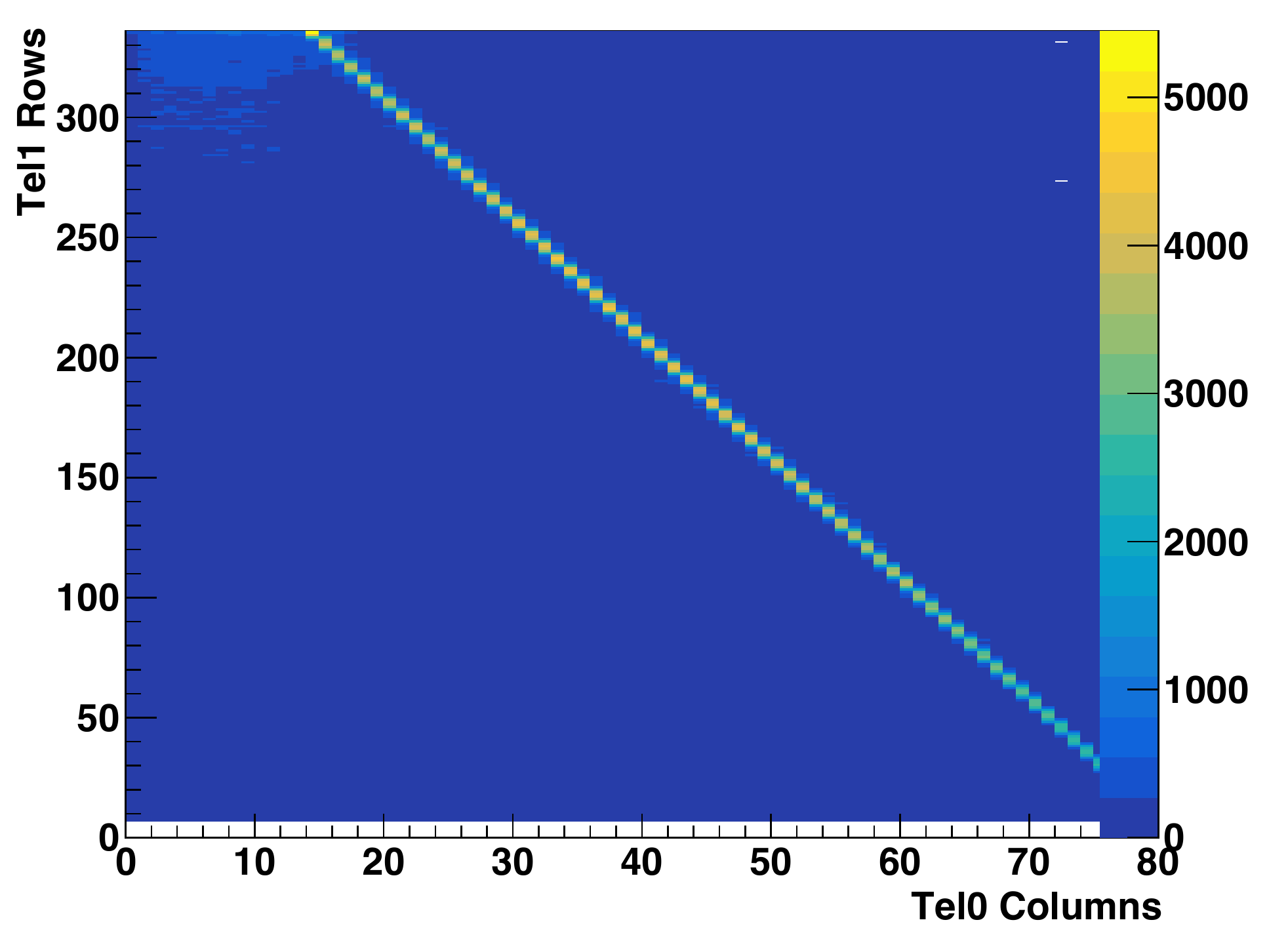}
         \caption{}
         \label{fig:cowb}
     \end{subfigure}
     \hfill
     \begin{subfigure}[b]{0.49\textwidth}
         \centering
         \includegraphics[width=\textwidth]{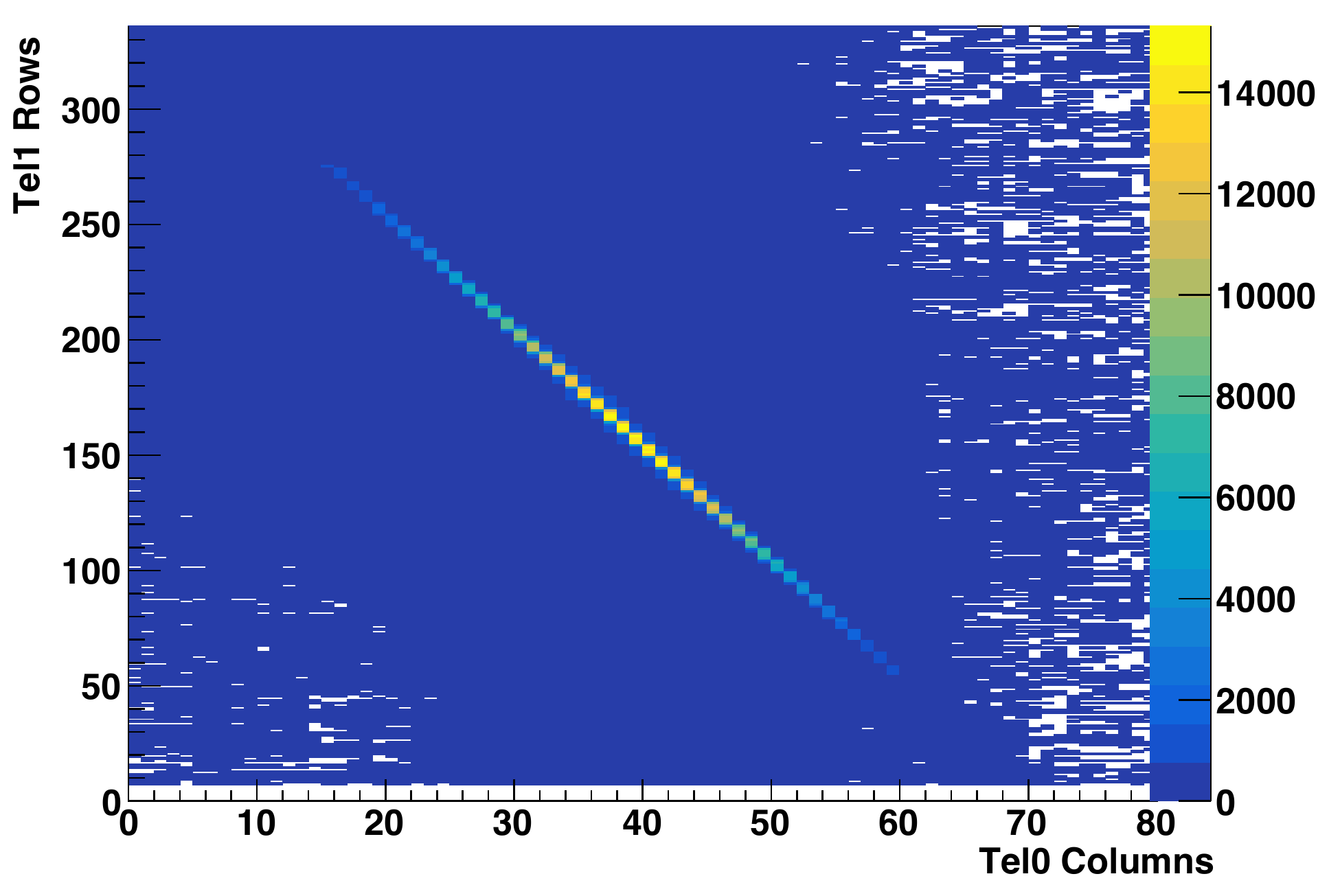}
         \caption{}
         \label{fig:conb}
     \end{subfigure}
    \caption{120 GeV beam energy profile with occupancy plots: (a) Wide beam profile (b) Narrow beam profile. Correlation between columns of the first plane and rows of the second plane: (c) For wide beam (d) For narrow beam}
    \label{fig:BP}
\end{figure}

The cosmicGui has an online monitoring feature that monitors the basic characteristic plots like hit occupancy, correlations, timing, and cluster size. The hit occupancy plots show the map of the number of pixels for a given sample of events. The alignment of the planes can be monitored using the correlation hit plots between the neighboring planes. The position between the planes is fixed during data taking. The first telescope plane's occupancy plots using wide and narrow beam settings are shown in Figures~\ref{fig:WB} and~\ref{fig:NB}, respectively. Correlation between the columns of the first plane and rows of the second plane are shown in Figures~\ref{fig:cowb} and~\ref{fig:conb}.

\newpage
\subsection{Cluster Size}
\begin{figure}[!h]
    \centering
    \includegraphics[width=0.55\textwidth]{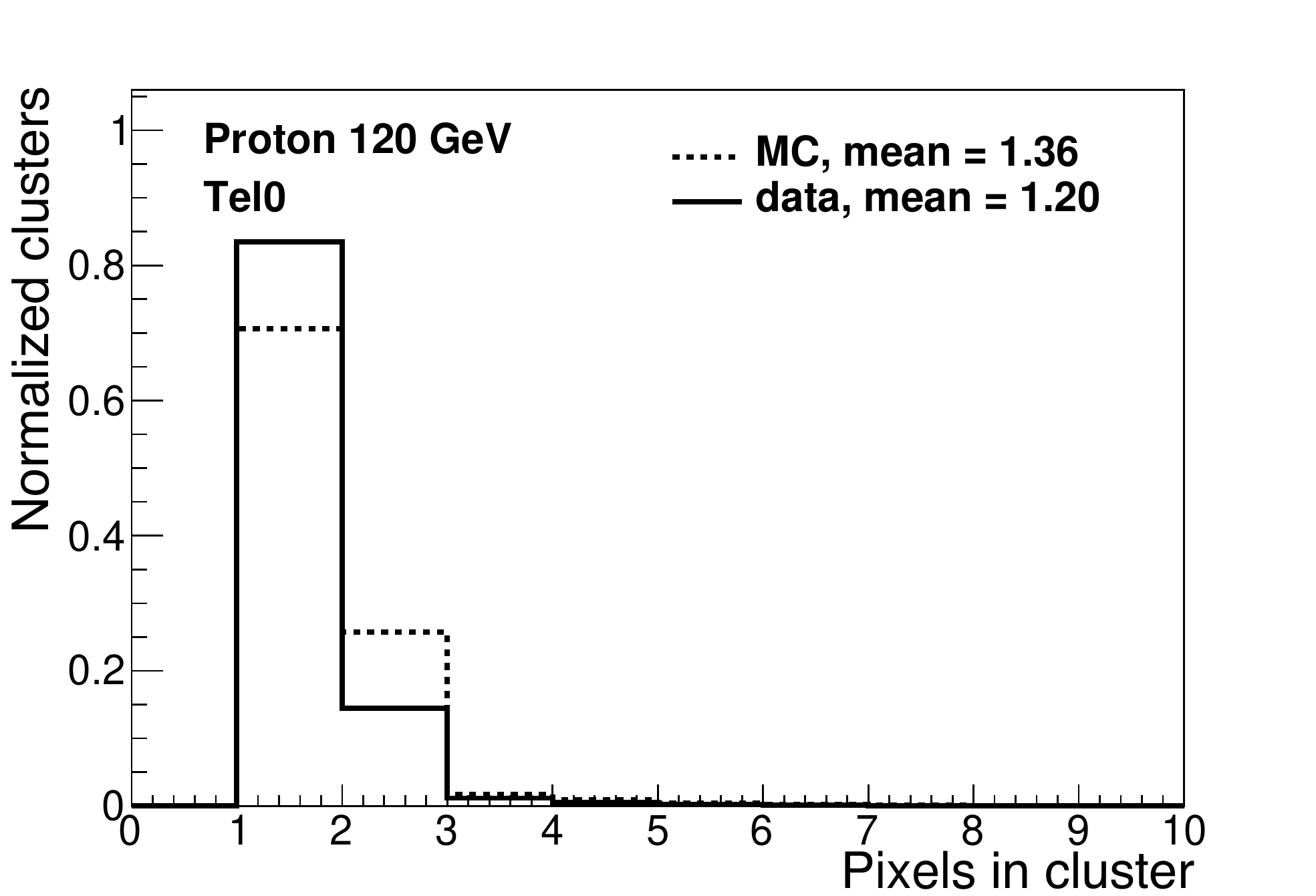}
    \caption{Normalized plot of cluster size in data and simulation for the first telescope plane}
    \label{fig:cs0x}
\end{figure}

When a particle passes through the tracking plane, it can fire more than one pixel per hit. The clustering algorithm groups such clusters together produced due to the single hit. In the current test beam setup, the observed average cluster size is larger than one. The simulation results also show a similar cluster size. Figure~\ref{fig:cs0x} shows the observed and simulated cluster sizes for the first telescope plane (Tel0). All six telescope planes have similar cluster-size responses. 

\subsection{Tracking Efficiency}
The tracking procedure in the Proteus software allows one to evaluate the tracking efficiency of each pixel within the tracking plane. The track must pass through all six tracking planes and should have a cluster in each plane. This condition helps to eliminate the tracks with larger scattering angles. If the extrapolated track matches the cluster in the given plane, then that track is included in the efficiency measurement.
\begin{figure}[!h]
    \centering
    \includegraphics[width=0.65\textwidth]{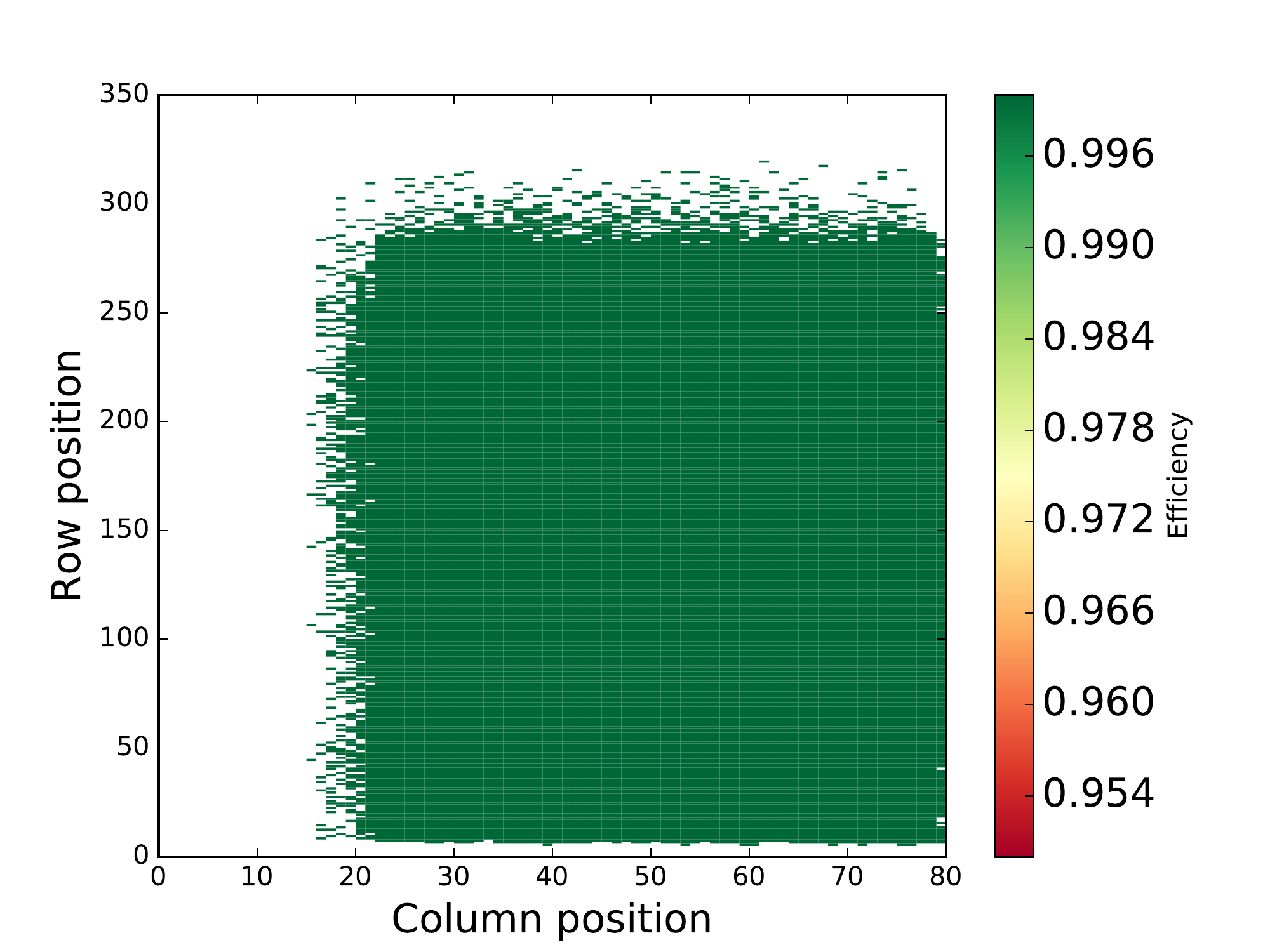}
    \caption{Track detection efficiency of the first telescope plane}
    \label{fig:trkeff}
\end{figure}

For each plane, one can set a range of included pixels within the framework. For the current setup in the test beam, each tracking plane's efficiency is measured by considering that plane as a DUT. In each plane, the pixel range is set for the efficiency measurement. The way chips in the telescope planes are aligned with each other, the cooling plates from other telescope planes slightly overlap with some rows and columns of the active chip. This effect reflects in the efficiency plot shown in Figure~\ref{fig:trkeff}. Therefore, only the active range of the rows and columns of the pixels are considered while calculating the track efficiency. The detection efficiency for each plane is around 99\%, as shown in Figure~\ref{fig:trkeff}.

\section{Spatial Resolution}
\label{sec:reso}
\subsection{Measured Residuals}
\begin{figure}[!h]
    \centering
    \begin{subfigure}[b]{0.49\textwidth}
    \centering
    \includegraphics[width=\textwidth]{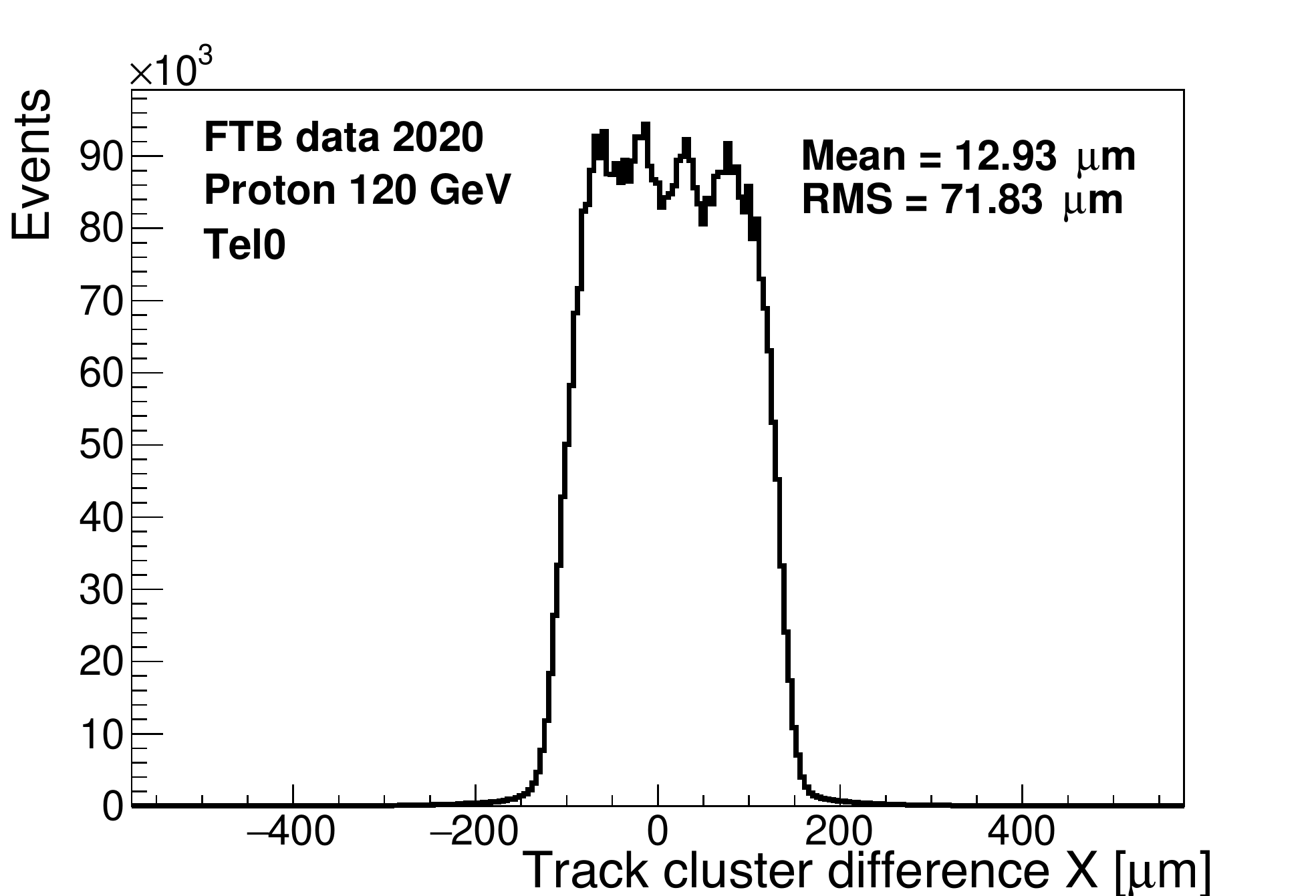}
    \caption{}
    \label{fig:res0x}
    \end{subfigure}
    \hfill
    \begin{subfigure}[b]{0.49\textwidth}
    \centering
    \includegraphics[width=\textwidth]{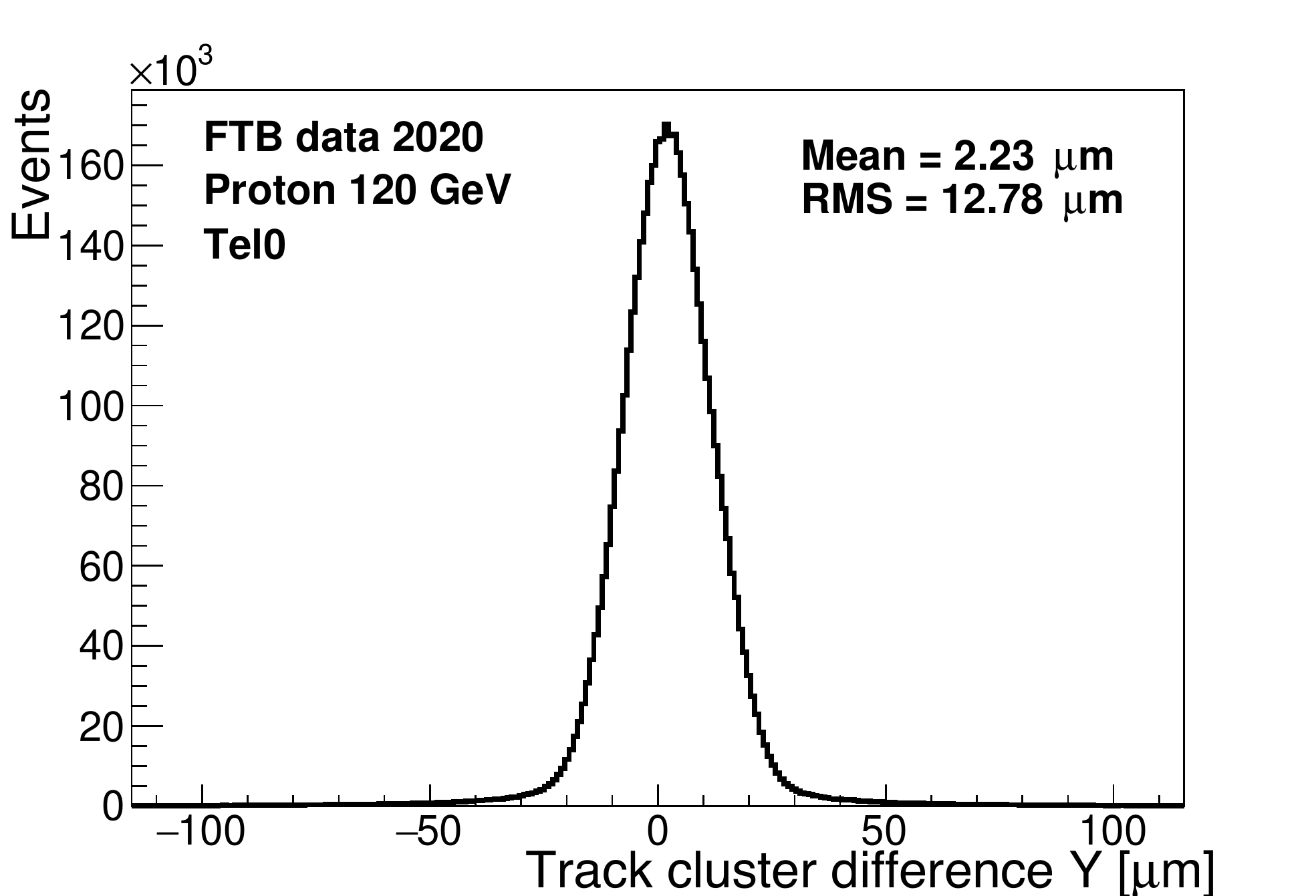}
    \caption{}
    \label{fig:res0y}
    \end{subfigure}
    \caption{Residual distribution in the first telescope plane after alignment (a) in the X-direction (b) in the Y-direction }
    \label{fig:measureres}
\end{figure}

APTT was commissioned during the test beam period in January-February 2020 at FTBF. The test beam data is reconstructed in three steps: clustering, alignment, and track reconstruction as described in Section~\ref{sec:reco}. Analysis shows that the tracks are reconstructed with 99\% efficiency. The track-fit parameter is used to derive the spatial resolution of the telescope. The digital hit information is taken into account during the reconstruction step; therefore, the resolution of the detector is mainly dependent on the position of the telescope planes and their pixel granularity.
Figure~\ref{fig:measureres} shows the residual distribution of the first telescope plane in the X and Y direction after alignment. The residual in the Y-direction follows the expected Gaussian distribution, whereas the X-residual exhibits the multiple peak structure. The non-square geometry of the pixels causes multiple peak structures in the X-residual. The detailed explanation of these phenomena is explained in Ref. \cite{Benoit_2016}.

Both the X and the Y residuals of the first telescope plane are centered at zero and the standard deviation of the distributions are $\sigma_{X,meas} =$ 71.83 $\mu$m and $\sigma_{Y,meas} =$ 12.78 $\mu$m, respectively.

For a binary readout the residuals in each direction can be calculated using the pitch of the pixels using the formula  $\sigma_{cal} =$ pitch /$\sqrt{12}$ \cite{rossi2006pixel}.  The pitch of the FE-I4B telescope plane in the X-direction is 250 $\mu$m, and therefore, the residual is $\sigma_{x,cal} =$ 72.2 $\mu$m. The pixels in Y-direction are 50 $\mu$m with corresponding residual width of $\sigma_{y,cal} =$ 14.4 $\mu$m. As shown in Figure~\ref{fig:cs0x}, clusters contained more than one hit, which leads to a better estimation of measured residual due to the charge sharing effect.
\subsection{Simulated Residuals}
\begin{figure}[!h]
    \centering
    \begin{subfigure}[b]{0.49\textwidth}
    \centering
    \includegraphics[width=\textwidth]{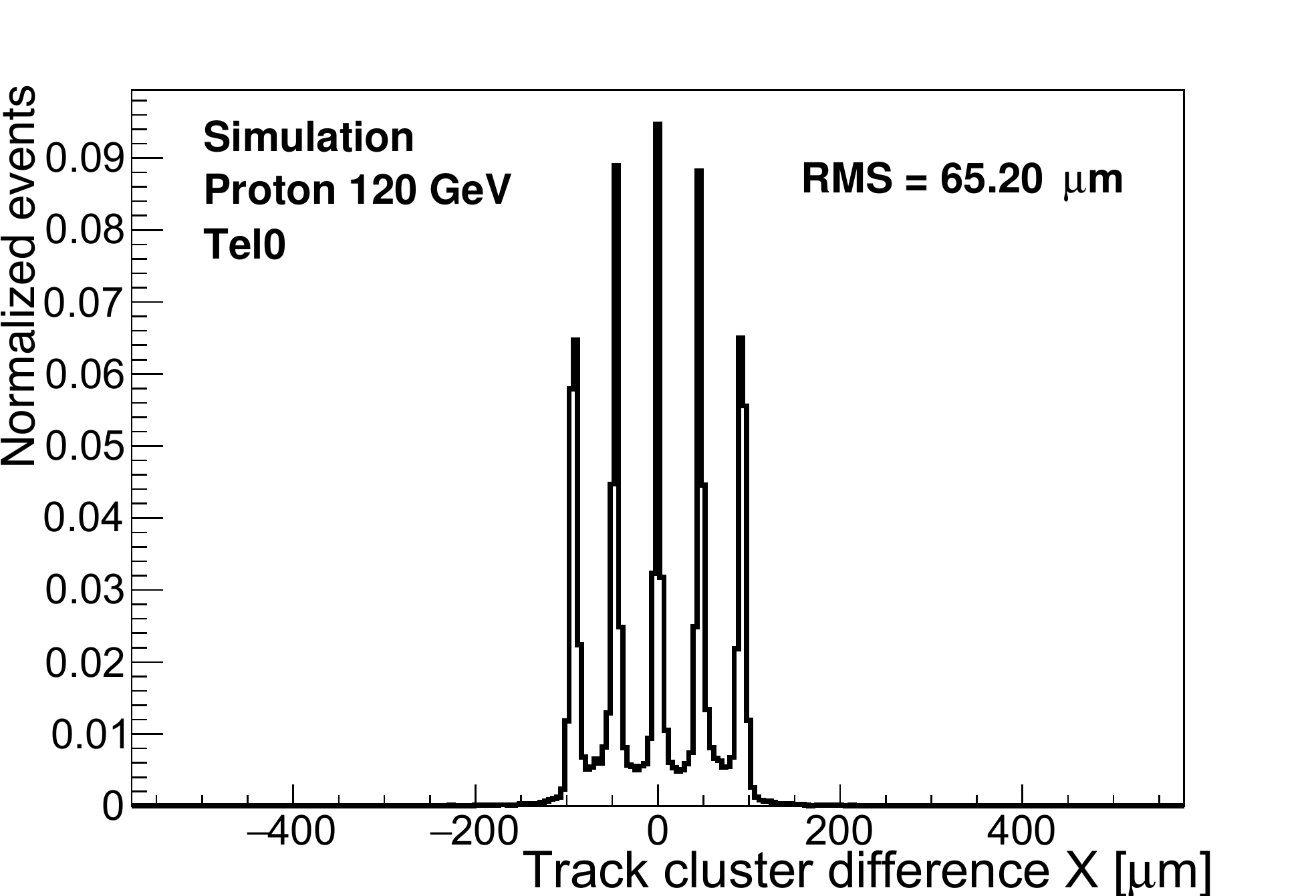}
    \caption{}
    \label{fig:sres0x}
    \end{subfigure}
    \hfill
    \begin{subfigure}[b]{0.49\textwidth}
    \centering
    \includegraphics[width=\textwidth]{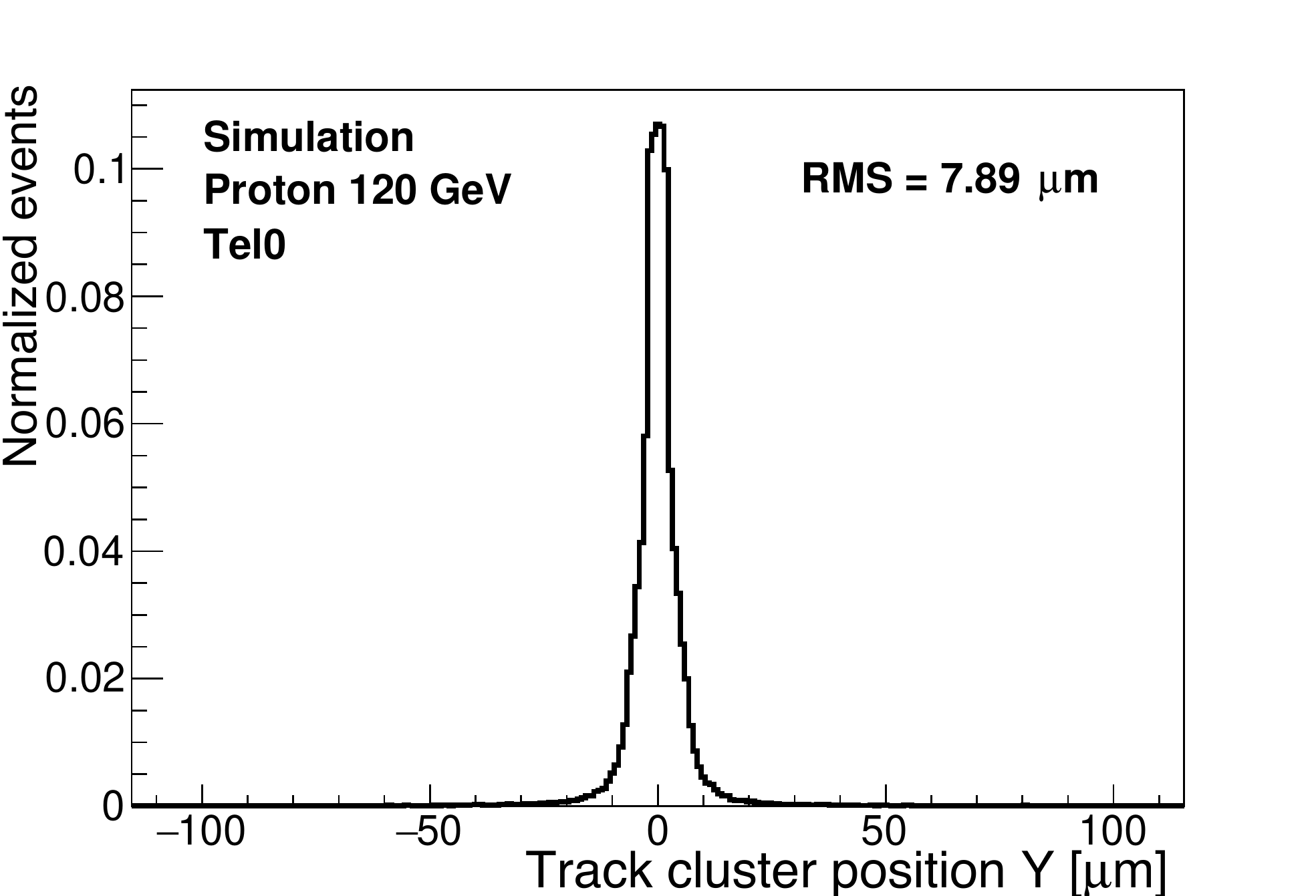}
    \caption{}
    \label{fig:sres0y}
    \end{subfigure}
    \caption{Simulated residual of the first telescope plane (a) in the X-direction (b) in the Y-direction }
    \label{fig:simres}
\end{figure}

The telescope performance is also studied using simulations based on the test beam scenario. The simulation is performed as described in Section~\ref{sec:simu}. The geometry and distance between the planes are set to the same values used in the test beam data analysis. The second and fifth planes in the beamline are rotated by 90$^{\circ}$ to achieve a better residual, as is the case in the actual test beam setup. The simulated residuals in the X and Y directions are shown in Figure~\ref{fig:simres}. The residual in the X direction is $\sigma_{x,sim} =$ 65.2 $\mu$m and in the Y-direction it is $\sigma_{y,sim} =$ 7.9 $\mu$m. 
The simulation residuals are smaller than the calculated residuals. This can be explained by the fact that the simulated clusters tend to contain more pixels, as shown in Figure~\ref{fig:cs0x}, resulting in a better resolution due to charge sharing. Details on the relationship between spatial resolution and multiple hit clusters is explained in Ref. \cite{rossi2006pixel}.

The simulated residual in the X-direction shows five peaks equally spaced by 50$\mu$m. These peaks are distinct compared to the observed residual as the simulated residuals are not affected by the detector's misalignment or inefficiencies. The rectangular pixel size of 250 $\times$ 50 $\mu$m$^{2}$ is solely responsible for this feature. Since two out of six planes are rotated by 90$^{\circ}$, they improve the granularity of the tracking in the X-direction. It also differentiates the tracks produced with the 50 $\mu$m resolutions, which leads to having the five possible reconstructed track positions. Therefore, instead of having the flat top of the residuals, it has a distinct five-peak structure.

\begin{figure}[!h]
    \centering
    \includegraphics[width=0.55\textwidth]{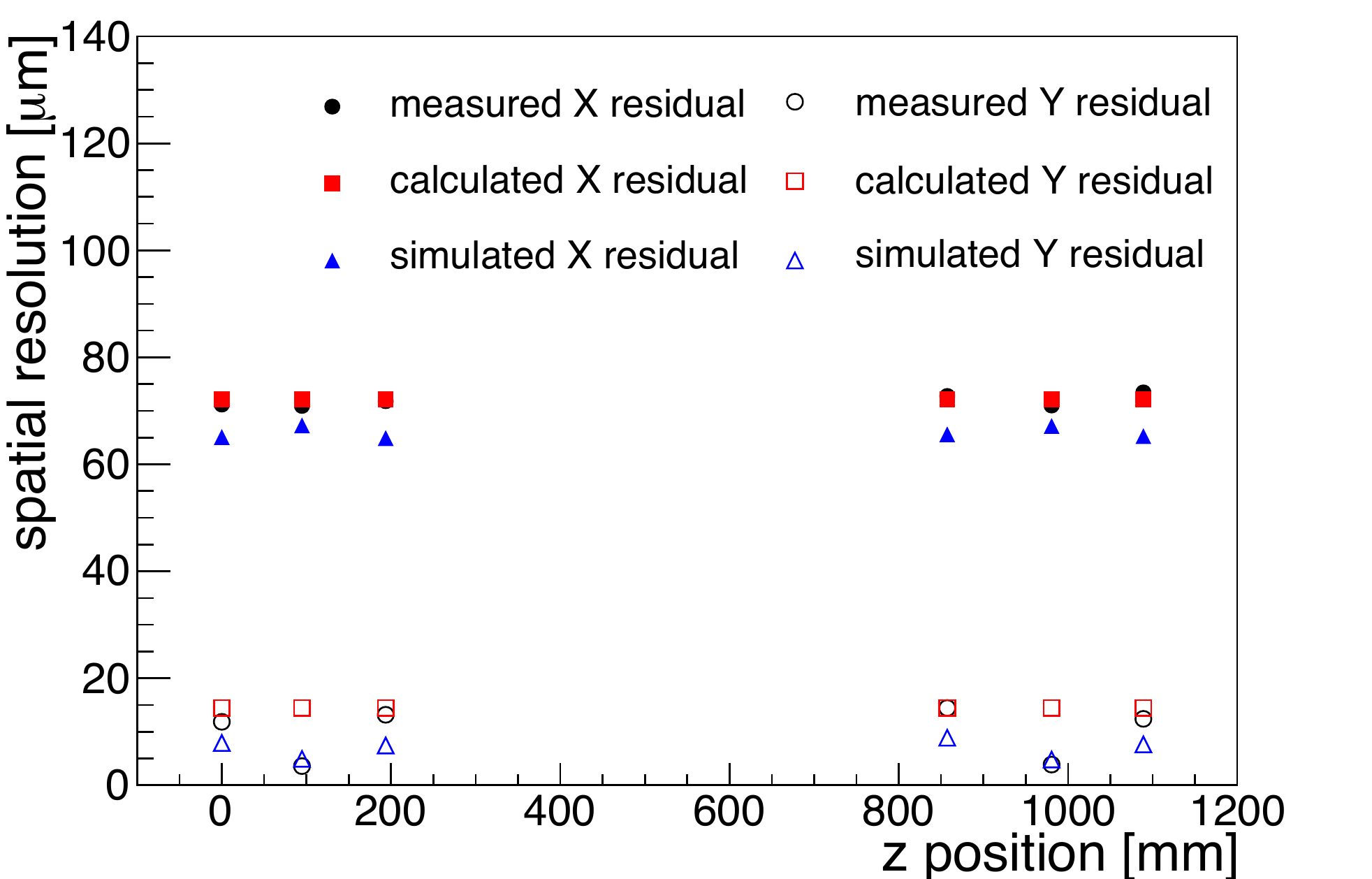}
    \caption{Measured, calculated and simulated residuals summary}
    \label{fig:sumres}
\end{figure}

To conclude, the spatial resolutions of six telescope planes measured at the test beam are within an agreement range with simulated and calculated residuals. Figure~\ref{fig:sumres} shows a summary of all these three types of residuals.

\section{Summary}\label{sec:summary}
The Argonne FE-I4B pixel tracking telescope is installed successfully at the Fermilab Test Beam Facility in the MT6.1B enclosure.  The telescope commissioning data were taken during the January-February 2020 test beam campaign. The telescope will be used to characterize and test various pixel detectors and DAQ systems developed for the HL-LHC ATLAS upgrade. The performance of the telescope has been studied and compared with simulated results. Characteristic plots like cluster size, pixel efficiency, and spatial resolution are made. The observed residual of the telescope plane in the X-direction is \mbox{$\sigma_{x,meas} =$ 71.83 $\mu$m} whereas the Y residual is \mbox{$\sigma_{y,meas} =$ 12.78 $\mu$m}. These residual values are consistent with expected residuals calculated using the pitch of the detector. The expected residual in X-direction is \mbox{$\sigma_{x,cal} =$ 72.2 $\mu$m,} whereas it is $\sigma_{y,cal} =$ 14.4 $\mu$m in Y-direction. The AllPix$^{2}$ software is used to simulate the test beam scenario. The simulated residuals are also within the experimental limits and they are \mbox{$\sigma_{x,sim} =$ 65.2 $\mu$m} in X-direction and \mbox{$\sigma_{y,sim} =$ 7.9 $\mu$m} in Y-direction. 

\acknowledgments 
The Argonne electronics group provided generous support to produce custom design PCB boards and specialized cables. Funding from Argonne National Laboratory, provided by the Director, Office of Science, of the U.S. Department of Energy under Contract No. DE-AC02-06CH11357.

Our sincere thanks to FTBF for providing the required infrastructure and their support. This document was prepared by Argonne National Laboratory (FTBF 1224) using the resources of the Fermi National Accelerator Laboratory (Fermilab), a U.S. Department of Energy, Office of Science, HEP User Facility. Fermilab is managed by Fermi Research Alliance, LLC (FRA), acting under Contract No. DE-AC02-07CH11359.

Our sincere thanks to Mathiew Benoit for helping us debug the DAQ issues and Moritz Klein for providing support for the Proteus software.
Our sincere thanks to the University of Oklahoma and Oklahoma State University for providing support during the test beam.

\bibliographystyle{JHEP}
\bibliography{ref}

\end{document}